\documentclass[aps,prl,twocolumn,superscriptaddress,showpacs,footinbib]{revtex4-1}
\usepackage{epsfig}
\usepackage{amsmath}
\usepackage{amssymb}
\usepackage{amsfonts}
\usepackage{graphicx}
\usepackage{alltt}
\usepackage{latexsym}

\def\sc {\scriptscriptstyle}
\def\spc {{\hskip  0.30mm}}
\def\spm {{\hskip -0.30mm}}

\def\Q   {{\mathcal Q}}
\newcommand{\I}{\mathcal{I}}

\newcommand{\tv}{\tau_{\spm \sc v}}
\def\G   {{\mathit \Gamma}}

\def\p   {{\text p}}

\begin{document}

\title{\bf Fundamental Diagram of Traffic Flow from Prigogine-Herman-Enskog Equation}

\author{W. Marques Jr.}
\affiliation{Departamento de F\'{\i}sica, Universidade Federal do Paran\'a, Caixa Postal 10944, 81531-990, Curitiba, Brazil}

\author{A. R. M\'endez}
\affiliation{Departamento de Matem\'aticas Aplicadas y Sistemas, Universidad Aut\'onoma Metropolitana - Cuajimalpa, 05348, Cuajimalpa, M\'exico}

\author{R. M. Velasco}
\affiliation{Departamento de F\'{\i}sica, Universidad Aut\'onoma Metropolitana - Iztapalapa, 09340, Iztapalapa, M\'exico}

\date{\today}

\begin{abstract}
\noindent Recent applications of a new methodology to measure fundamental traffic relations on freeways shows that many of the critical parameters 
of the flow-density and speed-spacing diagrams depend on vehicle length. In response to this fact, we present in this work a generalization of the Prigogine-Herman traffic 
equation for aggressive drivers which takes into account the fact that vehicles are not point-like objects but have an effective length. Our approach 
is similar to that introduced by Enskog for dense gases and provides the construction of fundamental diagrams which are in 
excellent agreement with empirical traffic data.    
\end{abstract}


\maketitle

Since the pioneering studies of Greenshields~\cite{Greenshields} at the beginning of the last century, fundamental diagram has attracted the attention 
of the scientific community because it gives a good insight of traffic conditions. For example, some interesting phenomena as 
traffic breakdown and jam propagation were addressed through the data obtained from flow-density fundamental diagrams \cite{Kernerlibro2}. As pointed out in the literature, 
there are many factors influencing the aspect of the fundamental diagram, e.g., road characteristics, vehicle composition, lighting conditions and wheather conditions. 
Despite all complexity, fundamental diagrams must satisfy the following basic requeriments \cite{Helbing01,Wageningen-Kessels}: (i) the mean vehicular speed $\bar{v}$ satisfy the condition $0 \leq \bar{v} \leq \bar{v}_{\spm \sc f}$, where $\bar{v}_{\spm \sc f}$ is the free flow speed; (ii) the traffic density $\rho$ satisfy the condition $0 \leq \rho \leq \rho_{\spm\sc j}$, where $\rho_{\spm \sc j}$ is the jam (or maximum) density; (iii) when the density goes to zero, the mean vehicular speed approaches the free flow velocity; (iv) at the jam density, vehicles stop moving 
so that the mean vehicular speed vanishes; (v) as the density increases, the mean vehicular speed decreases, i.e., $d\bar{v}/d\rho < 0$ for $0 < \rho \leq \rho_{\spm \sc j}$; 
(vi) the traffic flow $\Q=\rho \spc \bar{v}$ is strictly concave, so that $d^{\spc \sc 2}\Q/d\rho^{\sc 2} < 0$ and (vii) there exists a maximum flow $\Q_{\sc 0}$ which assures the existence of a critical density $\rho_{\spm\sc \text{c}}$ and a critical mean speed $\bar{v}_{\spm \sc \text{c}}$ such that $\Q_{\sc 0}=\rho_{\spm\sc \text{c}}\bar{v}_{\spm \sc \text{c}}$. It is important to mention that the critical density separates the fundamental diagram into two regions, namely: free flow region 
and congestion region. The free flow region is characterized by $\rho < \rho_{\spm \sc \text{c}}$ with $d\Q/d\rho > 0$, while the congestion region is defined by the conditions 
$\rho > \rho_{\spm \sc \text{c}}$ and $d\Q/d\rho < 0$.

Recently, Coifman \cite{Coifman2014,Coifman2015} introduces a new methodology to measure fundamental diagrams on freeways which eliminates the search for 
stationary conditions and provides a new level of accurancy to calibrate macroscopic and microscopic traffic models. After applying his method to three independent data set, 
Coifman found out that many of the critical parameters (e.g., the critical density and the slope of the congested regime $w=(d\Q/d\rho)_{\rho\spc =\rho_{\sc \text{c}}}$) appearing on the 
flow-density fundamental diagram depend on vehicle length. Thus, Coifman's results indicate that traffic flow models must be updated in order to take 
into account the length of the vehicles. In this letter, we address this problem by considering the derivation of the fundamental diagram from a 
mesoscopic point of view, i.e., we start from the gas-kinetic-like traffic equation \cite{Prigogine71,Iannini}
\begin{equation}\label{EQ1}
\frac{\partial f}{\partial t}+v\frac{\partial f}{\partial x}=-\frac{f-f_{\spm \sc 0}}{\tau}+\I(f,f),
\end{equation} 
where $f = f(x,v,t)$ denotes the distribution function. The distribution function is defined in such a way that $f(x,v,t)\spc dx\spc dv$ gives the 
number of vehicles at time $t$ with position around $x$ and velocity around $v$. The first term on the right-hand side of the kinetic traffic equation 
(\ref{EQ1}) describes the relaxation of the distribution function to a desired distribution 
$f_{\spm\sc 0}=f_{\spm\sc 0}(x,v,t)$ on a time scale $\tau$, while the second one is the so-called interaction term. By assuming that vehicles are not 
point-like objects but have a 
length $\ell$ and also require a safety distance $v \spc \tv$, where $\tv$ denotes drivers reaction time, it is possible to write the interaction term 
as
\begin{widetext}
\begin{equation}\label{EQ2}
\I(f,f)=\int_{\spm \sc v}^{\sc \infty}\!\!\! \chi \spc (1-\p)(v^\prime-v) f(x+s(v^\prime),v,t) f(x,v^\prime,t) \spc dv^\prime
-\int_{\spm \sc 0}^{\sc v}\!\! \chi \spc (1-\p)(v-v^\prime) f(x,v,t) f(x+s(v),v^\prime,t) \spc dv^\prime,
\end{equation}
\end{widetext}
where $s(v)=\ell + v\spc \tv$ represents the effective length of a vehicle moving with velocity $v$, $\p$ denotes the overtaking probability and $\chi$ is a factor 
describing the increase in the interaction rate due to the spatial extension of the vehicles. Following Prigogine and Herman \cite{Prigogine71}, we assume that the overtaking probability $\p$ and the relaxation time $\tau$ are given by $\p=1-z$ and $\tau=\tau_{\!\spc \sc 0}\spc (1-\p)/\p$, where $z=\rho/\spm \rho_{\spm \sc j}=\rho \ell$ is the so-called reduced density 
and $\tau_{\!\spc \sc 0}$ is a proportionality constant. The factor $\chi$ is given by the expression $\chi=(1-(z+\Q\spc \tv))^{\sc -1}$ which follows by taking into account the mean 
effective length $s(\bar{v})=\ell+\bar{v}\spc \tv$ common to two vehicles at the interaction process. Note that the factor $\chi$ is equal to unity 
for dilute traffic and increases with increasing density, becoming infinity when the density reaches the jam density \cite{Klar97}.

The construction of the fundamental diagram starts by considering the homogeneous and stationary solution of our gas-kinetic-like traffic equation, namely:
\begin{equation}\label{EQ3}
f=\frac{f_{\spm \sc 0}}{1-\gamma \bar{v}+ \gamma v},
\end{equation}
where $\gamma=\rho \chi (1-\p) \tau$. The above non-linear equation has two types of solutions corresponding to individual and collective 
flow patterns. Conditions determining the occurrence of both patterns were discussed in detail by Prigogine and Herman. To go further, let us consider 
the gamma distribution
\begin{equation}\label{EQ4}
f_{\spm \sc 0}=\frac{\alpha}{\G (\alpha)}\spc \frac{\rho}{\bar{v}_{\!\spc \sc 0}}\left(\frac{\alpha v}{\bar{v}_{\!\spc \sc 0}}\right)^{\!\!\alpha\sc -1}
\exp\left(-\frac{\alpha v}{\bar{v}_{\!\spc \sc 0}}\right),
\end{equation}  
where $\alpha > 1$ is the shape parameter and $\bar{v}_{\!\spc \sc 0}$ denotes the mean desired velocity. According to Velasco and Marques Jr. \cite{Velasco05}, 
the behavior of aggressive drivers can be described by a gamma distribution with a shape parameter which is directly connected with driver's aggressiveness. 
Note that the gamma distribution reduces to the exponential distribution when we set the shape parameter equal to one. The normalization condition for 
the homogeneous and stationary 
solution (\ref{EQ3}) can be written as 
\begin{equation}\label{EQ5}
\rho=\int_{\spm \sc 0}^{\sc \infty}\!\! \frac{f_{\spm \sc 0}}{\sigma+\gamma v}\spc dv,
\end{equation}
where $\sigma=1-\gamma \bar{v}$. Based on the expressions for $\chi$, $\p$ and $\tau$, we verify that $\sigma=1$ in the limit of vanishing density and 
may becomes negative at high density values. Since $\sigma$ cannot be negative, one can define a critical density $\rho_{\sc \text{c}}$ which 
determines the transition between individual and collective regimes. In the case of a gamma distribution, the critical density 
satisfies the condition 
 \begin{equation}\label{EQ6}
\beta z_{\spm \sc \text{c}}^{\sc 3}-\frac{\alpha}{\alpha-1}\spc (1-z_{\sc \text{c}})(1-z_{\sc \text{c}}-\Q_{\sc 0}\tv)=0, 
\end{equation}
where $\beta=\bar{v}_{\!\spc \sc 0}\tau_{\!\spc \sc 0}/\ell > 0 $ is a dimensionless parameter. 

In the collective regime, where the relation $\bar{v}\spm=\spm\gamma^{\sc -1}$ holds,    
traffic flow follows the collective flow curve
\begin{equation}\label{EQ7}
\Q\spm =\spm \Q_{\sc 0}\spc \frac{\displaystyle (1\spm -\spm z)^{\sc 2}}{\displaystyle (1\spm -\spm z) \Q_{\sc 0} \tv\spm 
+\spm (z/z_{\sc \text{c}})^{\sc 2}(1\spm -\spm z_{\sc \text{c}})(1\spm-\spm z_{\sc \text{c}}\spm -\spm \Q_{\sc 0} \tv)}
\end{equation}
which is independent of the shape parameter. This result reflects the fact that, at high densities, drivers abandon their 
individual characteristics and begin to follow the collective behavior of the system. Once critical traffic conditions are known, 
the slope $w=(d\Q/d\rho)_{\rho\spc =\rho_{\sc \text{c}}}$ of the 
fundamental diagram can be used to determine drivers reaction time $\tv$ in the collective regime. 

\begin{table}[ht]
\caption{Traffic and model parameters}
\vskip 2mm
\centering
\begin{tabular}{ccccccccc}
\hline                   
 & & 18-22\spc\spc \text{ft} & & 22-28\spc\spc \text{ft} & & 28-38\spc\spc \text{ft} & & 58-68\spc\spc \text{ft}\\ 
\hline\\[-3mm]
$\ell$\spc (\text{ft}) & & 20 & & 25 & & 33 & & 63\\[1mm]                        
$\Q_{\sc 0}$\spc (\text{veh/h}) & & 2350 & & 2120 & & 1525 & & 1170\\[1mm]
$\rho_{\sc \text{c}}$\spc (\text{veh/mi}) & & 48.4 & & 39.5 & & 35.1 & & 25.6\\[1mm]
$w$\spc ($\text{mi/h}$) & & -14.9 & & -16.6 & & -17.4 & & -26.9\\[1mm]
$\tau_{\spm \sc v}$\spc ($\text{s}$) & & 1.21 & & 1.34 & & 1.73 & & 1.99\\[2mm]
\hline
\end{tabular}
\end{table}

\begin{figure}[ht]
\vskip 0.65truecm
\includegraphics[width=0.375\textwidth]{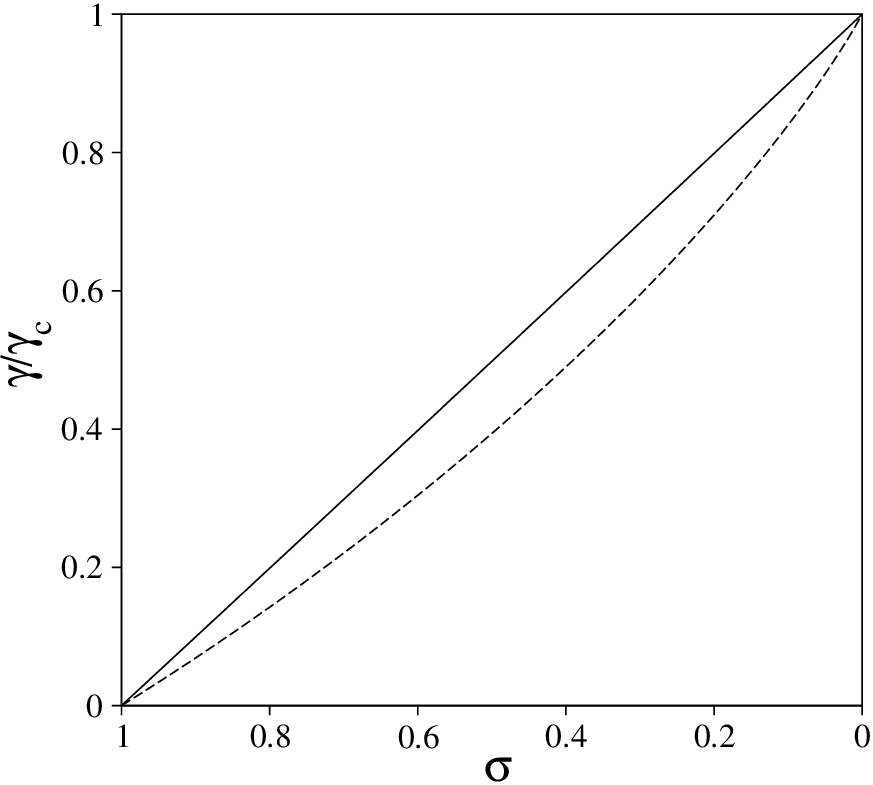}
\caption{\label{fig1} $\gamma/\gamma_{\sc \text{c}}$ as a function of $\sigma$ in the individual regime 
for the cases: $\alpha=120$ (solid line) and $\alpha=3$ (dashed line).}
\end{figure}

In the individual (or low-density) regime, we find that driver's aggressive behavior reaches its magnitude since traffic flow depends on the shape parameter. 
In this case, normalization condition (\ref{EQ5}) defines a function $\gamma (\sigma)$ which can be used to obtain the individual flow curve. Fig.~\ref{fig1}
shows the ratio $\gamma/\gamma_{\sc \text{c}}$ (where $\gamma_{\sc \text{c}}=\bar{v}_{\spm \sc 0}^{\sc -1} \alpha/(\alpha-1)$ denotes the value 
of $\gamma$ at the critical point) as a function of $\sigma$ for the gamma distribution. The curves displayed in this figure are for the cases: $\alpha=120$ (solid line) and $\alpha=3$ (dashed line). The linear behavior of the solid curve is quite evident and it allows us to write $\gamma = (1-\sigma)/\bar{v}_{\spm \sc 0}$ for the case $\alpha \gg 1$. Here, it is 
important to mention that the application of well-known solution methods of gas-kinetic theory to the Boltzmann-like traffic equation for aggressive 
drivers \cite{Marques13} allows to identify the inverse of the shape parameter as the prefactor of the velocity variance. Experimental data reported 
in the literature \cite{Helbing01} show that
the prefactor of the velocity variance is almost constant at the low-density region and they can be used to set values for the shape parameter. Thus, we can see that the case 
$\alpha=120$ is in agreement with empirical traffic data and traffic flow, in the individual regime, is given by the linear relation $\Q=\rho \spc \bar{v}_{\spm \sc 0}$.

\begin{figure}[t]
\vskip 1.1truecm
\includegraphics[width=0.425\textwidth]{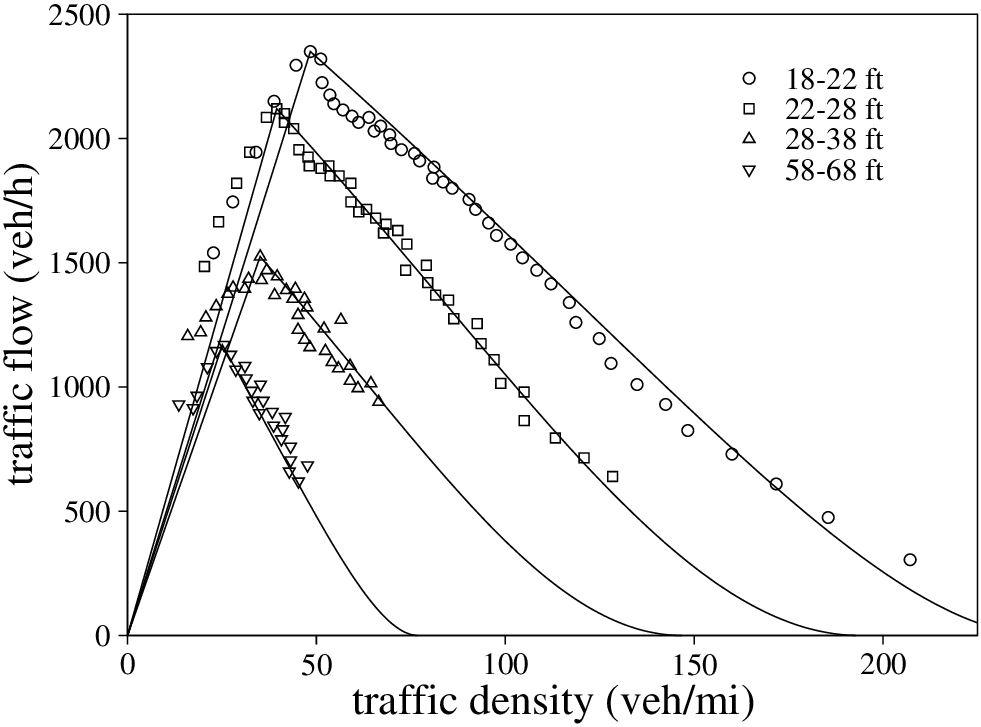}
\caption{\label{fig2} The flow-density curve for different vehicle lengths: solid lines represent the theoretical predictions derived from 
the non-local Prigogine-Herman traffic equation for a gamma distribution, while the dots reprensent the empirical traffic data of Coifman.}
\end{figure}

\begin{figure}[h]
\vskip 1.1truecm
\includegraphics[width=0.425\textwidth]{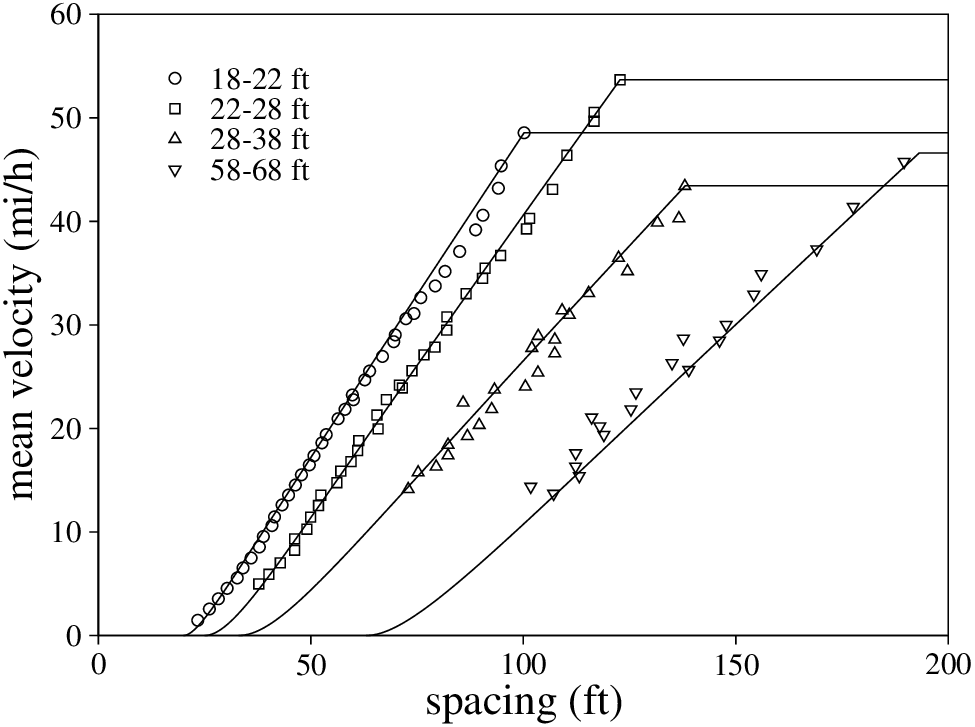}
\caption{\label{fig3} The speed-spacing curve for different vehicle lengths: solid lines represent the theoretical predictions derived from 
the non-local Prigogine-Herman traffic equation for a gamma distribution, while the dots reprensent the empirical traffic data of Coifman.}
\end{figure}

Figures \ref{fig2} and \ref{fig3} compare the theoretical flow-density and speed-spacing curves derived from our non-local Prigogine-Herman traffic 
equation with the empirical traffic data of Coifman for different vehicle length bins. Coifman's empirical traffic data \cite{Coifman2014,Coifman2015} 
follows by applying the single 
vehicle passage (svp) method to a primary data set which were collected from a two miles section of the I-80\break (in the East Bay portion of the San 
Francisco Bay area) by using Berkeley Highway Laboratory detector system. Regarding the theoretical results, they follow from our individual and collective flow curves by using traffic and model parameters given in Table I. In this table, the values of the critical density, flow capacity and slope of the flow-density curve in the collective regime were taken from Coifman's traffic data. Besides, we take the length of vehicles 
as mean length value in each length bin, while drivers reaction time in collective regime were obtained, as pointed out previously, from the slope of 
the flow-density diagram. 
One can verify that the agreement between experimental data and theoretical predictions is really impressive, even for long vehicles where curves show 
more scatter 
than for shorter vehicles. Such scatter event in empirical traffic data occur because there are fewer long vehicles on the highway than shorter ones.

In conclusion, we show in this letter that a non-local version of the Prigogine-Herman kinetic traffic equation for aggressive drivers can be used to describe properly the 
dependence of the fundamental diagrams on the length of the vehicles. Since such diagrams are central to most traffic 
flow models, we are quite sure that our mesoscopic\break description represents an important contribution to the understanding of some interesting phenomena as traffic breakdown and jam propagation. The present successful\break  comparison between theory and experiments marks a step towards a construction of a fundamental diagram for a multi-class traffic flow. 
In this case, modifications must be introduced in order to include different types of vehicles having different lengths and safety distances. For each vehicle class, 
one can assign a desired distribution function which,  together with the fractions of vehicles of each class, will govern the transition from individual to collective flow.

\subsection*{Acknowledgments}
The authors acknowledge support from CONACyT through grant number CB2015/251273.
\bibliography{TrafficPRL}
\bibliographystyle{unsrt}

\end{document}